\documentstyle[12pt]{article}

\catcode`\@=11

\font\twelvemsa=msam10 scaled 1200
\newcommand{\square}{\mbox{\twelvemsa \char'003}}

\font\twelvemsb=msbm10 scaled 1200

\font\sevenmsb=msbm7
\font\fivemsb=msbm5
\newfam\msbfam
\textfont\msbfam=\twelvemsb
\scriptfont\msbfam=\sevenmsb
\scriptscriptfont\msbfam=\fivemsb

\def\Bbb{\ifmmode\let\next\Bbb@\else
 \def\next{\errmessage{Use \string\Bbb\space only in math mode}}\fi\next}
\def\Bbb@#1{{\Bbb@@{#1}}}
\def\Bbb@@#1{\fam\msbfam#1}

\catcode`\@=12

\begin{document}

\begin{center}
{\Large\bf  Obstructions to Pin Structures}\\
\vspace*{0.3cm}
{\Large\bf on Kleinian Manifolds}\\
\end{center}
\vspace*{0.4cm}
\begin{center}
{\large L. J. Alty and A. Chamblin}\\
\end{center}
\vspace*{0.1cm}
\begin{center}
{\it Department of Applied Mathematics and Theoretical Physics,}\\
{\it Silver Street, Cambridge CB3 9EW, England}\\
\end{center}
\vspace*{0.5cm}

{\small We develop various topological notions on four-manifolds of Kleinian
signature $(- - + +)$. In particular, we extend the concept of `Kleinian
metric homotopy' [1] to non-orientable manifolds. We then derive the
topological obstructions to pin-Klein cobordism, for all of the pin groups.
Finally, we discuss various examples and applications which arise from this
work.}\\ \vspace*{0.6cm}

\begin{center}
{\bf I. ~Introduction}
\end{center}

Let $M$ be any smooth four-manifold,then we say that a metric $g$ on $M$ is
of {\it Kleinian signature} if it has signature $(- - + +)$. In recent work
[1], we derived the topological obstruction to spin-Klein cobordism, and in
this paper, we treat the interesting and non-trivial problem of extending
this work to non-orientable manifolds.

An orientable Kleinian manifold $(M, g)$ has orthonormal
frame bundle ${\tau}(M)$ with structure group $SO(2, 2)$. We say that $M$
admits a {\it spin structure} if and only if there exists a $2 - 1$ covering,
${\bar {\tau}}(M) ~{\longrightarrow}~ {\tau}(M)$, such that the following
diagram commutes:
\[
\begin{array}{cccccc}
{\mbox{Spin}}(2, 2) &{\longrightarrow} &{\bar {\tau}}(M)
&{\longrightarrow} & M & \\
 & & & & & \\
{\downarrow}{\mbox{\scriptsize 2 -- 1}} &
&{\downarrow}{\mbox{\scriptsize 2 -- 1}}  & &{\downarrow} &{\mbox{identity}} \\
 & & & & & \\
SO(2, 2) &{\longrightarrow} &{\tau}(M) &{\longrightarrow} & M &
\end{array}
\]
where ${\mbox{Spin}}(2, 2)$ is the double cover of $SO(2, 2)$.

When $M$ is non-orientable, one cannot reduce the tangent bundle ${\tau}(M)$ to
a bundle with structure group $SO(2, 2)$; indeed, ${\tau}(M)$ can only be
reduced to a bundle with structure group $O(2, 2)$. In analogy with the above
construction, we then seek all groups ${\bar O}(2, 2)$ which are double
covers of $O(2, 2)$; that is, we seek all groups ${\bar O}(2, 2)$ so that the
following sequence is exact:
\[
{1 ~{\longrightarrow}~ {\Bbb Z}_{2} ~{\longrightarrow}~ {\bar O}(2, 2)
{}~{\longrightarrow}~ O(2, 2) ~{\longrightarrow}~ 1}\; .
\]

In fact (see eg.\ [2,3]), there are eight distinct such groups which
are double covers of $O(2, 2)$. Following Dabrowski [2], we will call
these covers the {\it pin groups} for Kleinian signature and write them as
\[
h^{a, b, c}: ~{\mbox{Pin}}^{a, b, c}(2, 2) \longrightarrow O(2, 2)
\]
with $a, b, c ~{\in}~ {\{}+, -{\}}$.

In order to interpret the signs $a$, $b$ and $c$, it is convenient to keep
some of the terminology from Lorentzian geometry. Thus we will say that a
vector  $v~{\in}~ T_{p}(M)$ is {\it spacelike} if $g(v, v) > 0$,
{\it timelike} if $g(v, v) < 0$, and {\it null} if $g(v, v) = 0$.

Now we recall that $O(2, 2)$ is not path connected; there are four components,
given by the identity connected component, $O_{0}(2, 2)$, and the three
components corresponding to `space' inversion $S$, `time' inversion $T$, and
the combination of these two, $ST$ (i.e., $O(2, 2)$ decomposes into a
semi-direct product \footnote{ie. $O(2, 2)$ is the disjoint union
\[
{O(2, 2) ~{\simeq}~ O_{0}(2, 2) ~{\cup}~ S(O_{0}(2, 2)) ~{\cup}~ T(O_{0}(2, 2))
{}~{\cup}~ ST(O_{0}(2, 2))}\; .
\]
}, $O(2, 2) ~{\simeq}~ O_{0}(2, 2) ~{\odot}~ ({\Bbb Z}_{2} ~{\times}~ {\Bbb
Z}_{2})$). By `space' inversion, we mean reflection about a plane, $v^{\perp}$,
perpendicular to some spacelike vector $v$; likewise, `time' inversion is
reflection about a plane perpendicular to a timelike vector. The signs of $a,
b,$ and $c$ then correspond to the signs of the squares of the elements in
${\mbox{Pin}}^{a, b, c}(2, 2)$ which cover space inversion, time inversion,
and a combination of the two respectively.

Indeed, with these conventions we can write out the explicit form of the
groups ${\mbox{Pin}}^{a, b, c}(2, 2)$; they are given by the semi-direct
product [2]
\[
{{\mbox{Pin}}^{a, b, c}(2, 2) ~{\simeq}~ {\frac{({\mbox{Spin}}_{0}(2, 2)
{}~{\odot}~
C^{a, b, c})}{{\Bbb Z}_{2}}}}
\]
where the $C^{a, b, c}$ are the four double coverings of ${\Bbb Z}_{2}
{}~{\times}~ {\Bbb Z}_{2}$, as outlined in ([2], [3]).

On surveying the above constructions, one might wonder why we are concerned
with developing the obstruction theory for ${\mbox{Pin}}^{a, b, c}(2, 2)$
fibre bundles, since with ${\pi}_{1}(O(2, 2)) ~{\simeq}~ {\Bbb Z} ~{\times}~
{\Bbb Z}$, and so there is no way that the pin bundles will allow us to
represent all of the information contained in the tangent bundle in a
simply-connected manner [3]. Indeed, if we wished to represent the information
in ${\tau}(M)$ in a simply-connected manner, we would seek a bundle
${\xi}(M)$, with structure group ${\hat O}$ given by the exact sequence
\[
{1 ~{\longrightarrow}~ {\pi}_{1}(O(2, 2))
{}~{\simeq}~ {\Bbb Z} ~{\times}~ {\Bbb Z}~{\longrightarrow}~ {\hat O}
{}~{\longrightarrow}~ O(2, 2) ~{\longrightarrow}~ 1\; ,}
\]
wheras the pin groups are given by the short exact sequence
\[
{1 ~{\longrightarrow}~ {\Bbb Z}_{2} ~{\longrightarrow}~ {\mbox{Pin}}^{a, b,
c}(2, 2)
{}~{\longrightarrow}~ O(2, 2) ~{\longrightarrow}~ 1\; .}
\]
It follows that any pin bundle $P(M)$ (ie. any bundle with fibre
${\mbox{Pin}}^{a, b, c}(2, 2)$) will not represent information in a
simply-connected way. This means that at a point $p~{\in}~ M$ there exist
paths $\rho_{1}, \rho_{2} ~{\in}~ {\mbox{Pin}}^{a, b, c}(2, 2)$
which might act
on the fibre $P(M)|_{p}$ equivalently (in the sense that, for $x ~{\in}~
P(M)|_{p}, \rho_{1}(x) = \rho_{2}(x)$), but with the property that $\rho_{1}$
and $\rho_{2}$ (viewed as curves in ${\mbox{Pin}}^{a, b, c}(2, 2)$) are not
homotopic.  Indeed, one sees that the `particles' corresponding
to such a simply connected representation could have aribitrary
fractional statistics and would be `anyons' [4].  The point is that for both
Riemannian and
Lorentzian signature (in four dimensions) one obtains a simply connected
representation of tangent bundle information by passing to a fermionic (or pin)
bundle; it is only for Kleinian signature that this does not work and
one needs to introduce some anyonic structure.

At any rate, these mathematical considerations aside, the primary reason why we
wish to understand the obstructions to pin bundles comes from physics.
In particular,
recent work on signature change (see eg.\ [5--8]) has suggested that we should
allow for regions of non-Lorentzian signature in our description of nature. The
idea is that we should consider manifolds of the form $M ~{\cong}~ M_{L}
{}~{\cup}~ M_{R}$ and $M^{\prime} ~{\cong}~ M^{\prime}_{L} ~{\cup}~
M^{\prime}_{K}$, where (for example) $M_{R}$ is some Riemannian manifold,
$M_{L}$ and $M^{\prime}_{L}$ are some Lorentzian manifolds, and
$M^{\prime}_{K}$ is some Kleinian manifold (where ${\emptyset} ~{\neq}~
{\partial}M_{L} = {\Sigma} = {\partial}M_{R}$ and ${\partial}M^{\prime}_{L} =
{\Sigma}^{\prime} = {\partial}M^{\prime}_{K} ~{\neq}~ {\emptyset}$, so that
the signature is said to `change' across the three-surfaces ${\Sigma}$ and
${\Sigma}^{\prime}$, which are generically taken to be stationary with respect
to the ambient four-metrics). If one is going to assert that
there are `regions' of Kleinian signature, then one should try to make sense
of field theory [8] in signature $(- - + +)$. In particular, one must make
sense of the Dirac equation:
\begin{equation}
{i{\gamma}^{a}{\partial}_{a}{\psi} = 0}
\end{equation}
where ${\{}{\gamma}^{a}, {\gamma}^{b}{\}} = 2g^{ab}$. Solutions of (1) will
generically take values in some pin bundle, and so this is one reason the
cobordism
problem is so interesting. There are other physical applications for Kleinian
signature manifolds, including the $N = 2$ superstring theory [9]. In
this theory, the Weyl anomaly cancels provided the string
propagates in a four-dimensional target space, and if the worldsheet has
Lorentzian signature then the target space must have Kleinian signature.
\vspace*{0.6cm}

\begin{center}
{\bf II. ~Klein metric homotopy}
\end{center}

We wish to understand the topology of Klein metrics on four-manifolds which
are not necessarily orientable. The fundamental result we
begin with is the following lemma of Steenrod [10] \vspace*{0.3cm}

{\noindent {\bf Lemma 1.}~{\it Let $M$ be a smooth four-manifold without
boundary. Then $M$ admits a globally defined (non-singular) Klein metric if
and only if there exists a globally defined (non-singular) field of 2-planes
on $M$.}}\vspace*{0.3cm}

In [1] we restricted our consideration to fields of oriented 2-planes;
that is, since we were only considering orientable four-manifolds $M$, we
assumed that there were no closed loops, ${\gamma}$, in $M$ around which we
could propagate a 2-plane, $P$, and end up with the opposite orientation (of
the plane $P$). Technically, this meant that we assumed our plane fields to be
sections of the fibre bundle over $M$ with fibre $G_{2, 4} ~{\cong}~ S^{2}
{}~{\times}~ S^{2}$, where $G_{2, 4}$ is (by definition) the set of
oriented 2-planes in ${\Bbb R}^{4}$.

If, however $M$ is non-orientable then there
will exist loops in $M$ such that, when we propagate plane fields around them,
the orientation of the planes will be reversed as shown in Fig. 1).
In this case, we must now define a {\it plane field} to be a section of the
bundle of unoriented planes. That is, let ${\tilde G_{2, 4}}$ denote the
set of unoriented plane fields in ${\Bbb R}^{4}$. Then a field of
2-planes is a section of the fibre bundle with fibre ${\tilde G_{2, 4}}$.

This situation is reminiscent of what happens in Lorentzian geometry when one
passes from the study of time-orientable geometries to non-time-orientable
geometries ([3,11]); there, one passes from a vector field to a `line' field
(i.e., an undirected or unoriented vector field). However, the analogy should
not be taken too far. In Lorentzian geometry, non-time-orientability is a
serious matter since it implies that we have no local notion of an `arrow of
time', and thus many of our thermodynamical notions become tenuous.

For Kleinian geometry, there is no `arrow of time', since at
a point there is a `2-plane's' worth of timelike directions. In other words,
when we identify a Kleinian metric $g$ with a 2-plane field $P$, we can
essentially take $P$ to be the plane spanned by the set of timelike directions
(at each point). Indeed, in this paper we will always take $P$ to be a
`timelike' plane field. It follows that there is no sensible notion of causal
structure, or of causality, in a Kleinian manifold. In fact, there are `closed
timelike curves' through every point, as shown in Fig. 2. One can always just
`rotate' into one's own past. Indeed, `time' itself has a chirality (i.e., the
orientation of the plane field $P$). These considerations, if anything, make
it clear that orientability is less relevant in Kleinian geometry than
it is in Lorentzian geometry.

Now that we have made sense of what we mean by a `field of unoriented
2-planes', we need to consider the obstruction to constructing such a 2-plane
field which is globally non-singular. To do this, we first examine the
details of Hirzebruch and Hopf's [12] original treatment of the subject.

Let $M$ be an oriented smooth manifold, and $P$ some field of oriented
2-planes on $M$. Generically, $P$ will be singular on a finite set of points,
${\{}p_{1}, p_{2}, ..., p_{n}{\}}$ in $M$. Each singularity $p_{i}$ of $P$ will
have associated to it an index. The index of the singularity $p_{i}$ is
the homotopy type of the map (defined by the plane field $P$) from a little
three-sphere, $S^{3}(p_{i})$, surrounding $p_{i}$ to $G_{2, 4}$. Such homotopy
classes are in one-to-one correspondence with elements of ${\pi}_{3}(G_{2, 4})
{}~{\simeq}~ {\Bbb Z} ~{\oplus}~ {\Bbb Z}$. Thus, the index of $P$ at $p_{i}$
is
classified by a pair of integers (intuitively, this index measures the
`winding' of $P$ as one moves around $S^{3}(p_{i})$). We denote the index of
$P$ at $p_{i}$ by the symbol ${\mbox{ind}}(P, p_{i})$. Since there are
generically finitely many singular points $p_{i}$, one can form the
{\it index} of $P$ on $M$:
\[
{{\mbox{index of $P$ on
$M$}} = {\displaystyle\sum_{i = 1}^{n}} ~{\mbox{ind}}(P, p_{i})\; .}
\]

In [12] Hirzebruch and Hopf developed a result which gives the exact form of
the
index for orientable manifolds without boundary. The statement of their
result is as follows:

Let $M$ be an oriented compact four-manifold without boundary. Let $H$ denote
the free abelian group $H^{2}(M, {\Bbb Z})$/torsion subgroup, and let $S$
denote the intersection pairing on $H$ defined by the cup-product (ie. $S$
defines a map from $H ~{\otimes}~ H$ to ${\Bbb Z}$ by taking the cup-product
of elements in $H$ and evaluating them on the fundamental orientation class of
$M$). Define the coset $W ~{\subseteq}~ H/2H$ by $w ~{\in}~ W$ if $S(w, x) =
S(x, x) ~{\rm mod~2}$ for all $x ~{\in}~ H$. Finally, let ${\Omega}$ denote
the set of integers ${\{}S(w, w)|w ~{\in}~ W{\}}$, then we have [12,13]
\vspace*{0.3cm}

{\noindent {\bf Theorem 1.} {\it Let $M$ be an oriented compact
four-manifold without boundary. Then $M$ has a field of 2-planes with finite
singularities. The total index of such a field is given by a pair of integers
$(a, b)$. The following integers and only these, occur as the index for some
plane field on $M$:} \[
{a = {\mbox{$\frac{1}{4}$}}({\alpha} - 3{\sigma} - 2{\chi}),\; b=
{\mbox{$\frac{1}{4}$}}({\beta} - 3{\sigma} + 2{\chi})}
\]
{\it where ${\alpha}, {\beta} ~{\in}~ {\Omega}, {\chi} = {\chi}(M)$ denotes the
Euler
number of $M$, and ${\sigma} = {\sigma}(M)$ denotes the Hirzebruch signature of
$M$.}}\vspace*{0.3cm}

We wish to be able to calculate the index of a plane field on a non-orientable
manifold, and thus it is instructive to examine the proof of this
theorem to see exactly which steps are invalid when one passes to the
non-orientable case.

To begin with, recall that the {\it Stiefel manifold}, $V_{2, 4}$, is defined
to be the set of oriented dyads{\footnote{ie. a dyad is a pair of
vectors.}} ${\{}v_{1}, v_{2}{\}}$ in ${\Bbb R}^{4}$. It is clear that any dyad
${\{}v_{1}, v_{2}{\}}$ induces a plane $P$ (ie. $P$ is spanned by $v_{1}$
and $v_{2}$) and so we have the inclusion
\[
{{\varphi}: ~V_{2, 4} ~{\longrightarrow}~ G_{2, 4}}\; .
\]
Likewise, we can consider the Stiefel
manifold ${\tilde V_{2, 4}}$ of unoriented dyads in ${\Bbb R}^{4}$; then we
have the
inclusion
\[
{{\tilde {\varphi}}: ~{\tilde V_{2, 4}} ~{\longrightarrow}~ {\tilde G_{2,
4}}}\; .
\]

Now, in [12] the construction begins by considering the {\it skeleton} [14] of
$M$. Let $M^{1}$ denote the 1-skeleton, $M^{2}$ the 2-skeleton, etc., then
we always can put a dyad field ${\{}v_{1}, v_{2}{\}}$ (a section of the fibre
bundle with fibre $V_{2, 4}$ or ${\tilde V_{2, 4}}$) on the 2-skeleton. If $M$
is oriented, we can take this dyad field to be oriented; if $M$ is not
oriented, we will generically have to take the dyad field to be unoriented. We
then want to extend the dyad field on $M^{2}$ to a dyad field on $M^{3}$. In
the oriented case, the obstruction to doing this is $w_{3}(M)$ [14], the third
Stiefel-Whitney class of $M$. Of course, for a compact oriented $M$, $w_{3}(M)
= 0$, and so one is able to conclude [12] that the obstruction to extending a
plane field to all of $M$ must be an element of $H^{4}(M; {\pi}_{3}(G_{2, 4}))
{}~{\simeq}~ H^{4}(M; {\Bbb Z} ~{\oplus}~ {\Bbb Z}) ~{\simeq}~ H^{4}(M; {\Bbb
Z}) ~{\oplus}~ H^{4}(M; {\Bbb Z}) ~{\simeq}~ {\Bbb Z} ~{\oplus}~ {\Bbb Z}$.

When $M$ is not oriented, it might be thought that there is some other
obstruction to extending a dyad field (and thus a plane field) from $M^{2}$ to
$M^{3}$. However, this is not the case. Although Wu's formula [14] shows that
$w_{3}(M)$ can be non-vanishing for $M$ non-oriented, $w_{3}(M)$ will no
longer be the obstruction to extending the field $P$ to $M^{3}$ since we
are now allowing the plane field to be unoriented. Indeed, the obstruction to
this extension vanishes as long as $P$ is a section of a ${\tilde G_{2, 4}}$
bundle. It follows that the obstruction to extending an unoriented plane
field $P$ over an unoriented four-manifold $M$ is an element of $H^{4}(M;
{\pi}_{3}({\tilde G_{2, 4}})$). But $G_{2, 4}$ is the $2 - 1$ cover of
${\tilde G_{2, 4}}$ and so
\[
{{\pi}_{3}(G_{2, 4}) ~{\simeq}~ {\pi}_{3}({\tilde G_{2, 4}}) ~{\simeq}~ {\Bbb
Z}
{}~{\oplus}~ {\Bbb Z}}
\]
Thus, in both the oriented and non-oriented cases, the obstruction to extending
$P$ over $M$ is an element of $H^{4}(M; {\Bbb Z}$ $~{\oplus}~ {\Bbb Z})$.

At first, this may seem strange, since for a non-orientable manifold $M$,
$H^{4}(M; {\Bbb Z}) ~{\simeq}{\Bbb Z}_{2}$, and so the total index would seem
to be an element of
\[
{H^{4}(M; {\Bbb Z} ~{\oplus}~ {\Bbb Z}) ~{\simeq}~ H^{4}(M; {\Bbb Z})
{}~{\oplus}~
H^{4}(M; {\Bbb Z}) ~{\simeq}~ {\Bbb Z}_{2} ~{\oplus}~ {\Bbb Z}_{2}}\; ,
\]
and so the index is only defined `up to parity'.
However, as we shall see,
the parity of the index is the only thing relevant in the construction of our
obstructions.

Before continuing with the derivation of the form of the index,
it is necessary to recall some elementary topological
objects which we will make use of later.
To begin with, we have that a manifold $M$ admits a globally defined
metric of Kleinian signature if and only if the tangent bundle of $M$
(${\tau}(M)$) can be globally decomposed into the Whitney sum
\[
{{\tau}(M) ~{\simeq}~ {\tau}^{+} ~{\oplus}~ {\tau}^{-}}
\]
where ${\tau}^{+}$ is the subbundle of ${\tau}(M)$ generated by `spacelike'
vectors and ${\tau}^{-}$ is the subbundle generated by `timelike' vectors. Let
$w_{1}(M) = w_{1}({\tau}(M))$ denote the {\it first Stiefel-Whitney class} of
$M$. As is well known [14], $w_{1}(M) = 0$ if and only if $M$ is orientable.
Since $w_{1}$ is a 1-cochain, this means that $M$ is
orientable if and only if there are no closed loops, ${\gamma} ~{\in}~ M$,
such that $w_{1}[{\gamma}] ~{\neq}~ 0$. Under the Whitney sum,
$w_{1}({\tau}(M))$ can be decomposed as
\begin{equation}
{w_{1}({\tau}(M)) = w_{1}({\tau}^{+}) ~+~ w_{1}({\tau}^{-})} \; .
\end{equation}
We shall adopt the notation $w_{1}^{+} = w_{1}({\tau}^{+})$ and
$w_{1}({\tau}^{-}) = w_{1}^{-}$. Thus, $M$ is {\it space-orientable} if and
only if $w_{1}^{+} = 0$, and {\it time-orientable} if and only if $w_{1}^{-} =
0$. Note that if there exists some loop, ${\gamma} ~{\in}~ M$, such that $M$ is
neither space nor time-orientable, then $M$ is orientable since
$w_{1}({\tau}(M)) = w_{1}^{+} ~+~ w_{1}^{-} = 1 ~+~ 1 = 0\;{\rm mod~2}$. (We
always count mod 2 since these cochains always take values in ${\Bbb Z}_{2}$
[3]). As we shall see, $w_{1}^{+}$ and $w_{1}^{-}$ are critical components of
the obstructions to all of the pin structures.

Let us denote the {\it second Stiefel-Whitney class}, $w_{2}(M) =
w_{2}({\tau}(M))$. Recall that this class vanishes if and only if $M$ admits a
spin structure.

Now as in [3], we can apply Wu's formula and obtain the identity:
\begin{equation}
{(w_{2}(M) ~+~ w_{1}(M) ~{\smile}~ w_{1}(M)) ~{\smile}~ x_{2} =
x_{2} ~{\smile}~ x_{2}}
\;\;\;\;{\rm for\; any} x_{2} ~{\in}~H^{2}(M; {\Bbb Z}_{2}) \; ,
\end{equation}
where `${\smile}$' is the cup product [14]. Since we are allowing $M$ to be
non-orientable, we work in ${\Bbb Z}_{2}$-coefficients and write the
intersection
pairing
\begin{equation}
{h: ~H_{2}(M; {\Bbb Z}_{2}) ~{\times}~ H_{2}(M; {\Bbb Z}_{2})
{}~{\longrightarrow}~
{\Bbb Z}_{2}}\; .
\end{equation}
This is defined by $h(x, y) = x ~{ \bf {\cdot}}~ y = (x_{2} ~{\smile}~ y_{2})
{}~{\frown}~ w_{1}$, where $x_{2}, y_{2} ~{\in}~ H^{2}(M; {\Bbb Z}_{2})$
satisfy
$x_{2} ~{\frown}~ w = x$ and $y_{2} ~{\frown}~ w = y$, $w ~{\in}~ H^{4}(M;
{\Bbb Z}_{2})$ is the fundamental homology class and `${\frown}$' denotes cap
product. Taking the dual of equation (4) yields the intersection pairing on
$H^{2}(M; {\Bbb Z}_{2})$.

As we saw in [3], the following result of Kervaire and Milnor [15] holds even
for non-orientable $M$: \vspace*{0.3cm}

{\noindent {\bf Lemma 2.}~{\it Let $M$ be a
smooth four-dimensional manifold. Let $u({\partial}M)$ (the mod
2 Kervaire semi-characteristic) be given by}
\[
{u({\partial}M) = {\it dim}_{{\Bbb Z}_{2}}(H_{0}({\partial}M; {\Bbb Z}_{2})
{}~{\oplus}~
H_{1}({\partial}M; {\Bbb Z}_{2}))~{\rm mod~2}}
\]
{\it Then the rank of the intersection pairing, $h$, satisfies}
\[
{{\mbox{\it rank}}(h) = (u({\partial}M) ~+~ e(M))~{\rm mod~2}}
\]
{\it where $e(M)$ is the Euler number of $M$.}
\vspace*{0.3cm}

It is now easy to check [3] that ${\mbox{rank}}(h) = 0$ if and only if
\[
{w_{2} ~+~ w_{1} ~{\smile}~ w_{1} = 0 ~~{\rm mod~2}.}
\]
Combining this observation with Lemma 2 then yields
\vspace*{0.3cm}

{\noindent {\bf Lemma 3.} {\it Let $M$ be a smooth four-dimensional manifold
with tangent bundle ${\tau}(M)$. Then}}
\[
{w_{2}({\tau}(M)) ~+~ w_{1}({\tau}(M)) ~{\smile}~ w_{1}({\tau}(M)) = 0}
\]
\[
{{\Longleftrightarrow}}
\]
\[
{(u({\partial}M) ~+~ e(M))~{\rm mod~2} = 0}
\]
\vspace*{0.3cm}

We have now developed enough mathematical machinery to calculate the index of a
non-orientable plane field on a general (not necessarily oriented)
four-dimensional manifold $M$. Our basic strategy is the following: If $M$ is
oriented, we are done (we just apply the Hirzebruch-Hopf result, Theorem 1).
If $M$ is not orientable, we pass to the {\it oriented double cover}, ${\tilde
M}$, of $M$ and apply Theorem 1 on ${\tilde M}$. We then `push down' the plane
field ${\tilde P}$ on ${\tilde M}$, under the projection ${\pi}: ~{\tilde M}
{}~{\longrightarrow}~ M$, and deduce the form of the index of $P =
{\pi}^{*}({\tilde P})$ on $M$. Since any plane field $P$ on $M$ can be so
obtained, we thus derive the general form of the index on $M$.

Suppose then we are given some smooth four-dimensional manifold $M$ without
boundary, with Klein metric `$g_{K}$' defined on $M$. As we have seen, the
metric corresponds to some two-plane field $P$ on $M$. The singularities of the
metric $g_{K}$ therefore correspond to the singularities of the plane field
$P$. Construct the (oriented) $2 - 1$ cover over $M$, denoted ${\tilde
M}$, with projection ${\pi}:~ {\tilde M} ~{\longrightarrow}~ M$. Now
lift the plane field $P$ (which will generally be a section of a ${\tilde
G_{2, 4}}$ bundle, $B$, over $M$) to a plane field ${\tilde P}$ over
${\tilde M}$ (where ${\tilde P}$ will now be a section of a {\it $G_{2, 4}$}
bundle, ${\tilde B}$, over ${\tilde M}$).  Since
${\tilde M}$ is oriented, we know the form of the total index of ${\tilde P}$
on ${\tilde M}$ is, by Theorem 1,
\begin{equation}
{{\mbox{ind}}({\tilde P}, {\tilde M}) = {\mbox{$\frac{1}{4}$}}({\alpha} -
3{\sigma} - 2{\chi}, {\beta} - 3{\sigma} + 2{\chi})}
\end{equation}
{where ${\sigma} = {\sigma}({\tilde M}), {\chi} = {\chi}({\tilde M})$.
As in [1], we find that}
\begin{eqnarray}
{\alpha} = {\sigma}({\tilde M})~{\rm mod~8}\; , \nonumber \\
                                          \\
{\beta} = {\sigma}({\tilde M})~{\rm mod~8}\; , \nonumber
\end{eqnarray}
{hence}
\begin{eqnarray}
{\alpha} - {\sigma} = 8n, ~n ~{\in}~ {\Bbb Z}\; , \nonumber \\
                                                \\
{\beta} - {\sigma} = 8m, ~m ~{\in}~ {\Bbb Z}\; , \nonumber
\end{eqnarray}
so the index becomes
\begin{equation}
{{\mbox{ind}}({\tilde P}, {\tilde M}) = {\mbox{$\frac{1}{4}$}}\bigl(8n -
2({\sigma}({\tilde M}) ~+~ {\chi}({\tilde M})),\; 8m - 2({\sigma}({\tilde M}) -
{\chi}({\tilde M}))\bigr)}
\end{equation}
where $m$ and $n$ are some integers. It follows that we must
have
\begin{equation}
{{\sigma}({\tilde M}) = {\chi}({\tilde M})~{\rm mod~2}}\; .
\end{equation}
We must now determine how the parity of ${\chi}({\tilde M})$ is related
to the parity of ${\chi}(M)$. To do this we
introduce a new invariant: \vspace*{0.2cm}

{\noindent {\bf Definition:}~ {\it Let $M$ be a smooth
four-dimensional manifold with boundary ${\partial}M ~{\cong}~ {\Sigma}_{1}
{}~{\cup}~ {\Sigma}_{2} ~{\cup}~ ...  ~{\cup}~ {\Sigma}_{n}$ the disjoint union
of finitely many (not necessarily orientable) closed three-manifolds. Let
${\partial}{\tilde M} ~{\cong}~ {\tilde {\Sigma}_{1}} ~{\cup}~ {\tilde
{\Sigma}_{2}} ~{\cup}~ ... ~{\cup}~ {\tilde {\Sigma}_{n}}$ denote the oriented
double cover{\footnote{That is, ${\tilde {\Sigma}_{i}}$ is the oriented double
cover of ${\Sigma}_{i}$ for each $i$, and ${\partial}{\tilde M}$ is the
disjoint union of the ${\tilde {\Sigma}_{i}}$.}} of ${\partial}M$. Then we
define the element $U({\partial}M) ~{\in}~ {\Bbb Z}_{2}$ by the formula}}
\[
{U({\partial}M) = (u({\partial}M) - u({\partial}{\tilde M}))~{\rm mod~2}}
\]

Thus, $U({\partial}M)$ measures (modulo 2) the total number of torsion
generators of $H_{1}({\partial}M)$ which are `destroyed' when we pass to the
double cover.

For example, suppose that ${\partial}M ~{\cong}~ S^{1}
{}~{\times}~ {\Bbb R}{\Bbb P}^{2}$, then $u({\partial}M) = 1$ since ${\Bbb
R}{\Bbb P}^{2}$ has a torsion generator; and ${\partial}{\tilde M}
{}~{\cong}~ S^{1} ~{\times}~ S^{2}$ so $u({\partial}{\tilde M}) = 0$, hence
$U(\partial M) = 1$. Similarly all the torsion generators are destroyed
if we take ${\partial}M ~{\cong}~ {\Bbb R}{\Bbb P}^{3}$.

On the other hand, there are torsion generators which do not completely
`unwrap'. For example, take ${\partial}M ~{\cong}~ S^{1} ~{\times}~ K$, where
$K$ denotes the Klein bottle. We then find that $u({\partial}M) = 0$, where
one of the $H_{1}({\partial}M)$ factors is torsion, and that
${\partial}{\tilde M} ~{\cong}~ S^{1} ~{\times}~ T^{2}$ has
$u({\partial}{\tilde M}) = 0$. Hence $U({\partial}M) = 0$, which makes sense
since the torsion generator in $K$ lifts to a non-trivial loop in $T^{2}$.

We must also introduce another new invariant, which we define as follows:
\vspace*{0.2cm}

{\noindent {\bf Definition:}~ {\it Let $M$ be a smooth four-manifold, with or
without
boundary. Then we define the element ${\delta}(M) ~{\in}~ {\Bbb Z}_{2}$ as
follows:}}
\[
{{\delta}(M) = \left\{ \begin{array}{cl}
0 &~~{\mbox{\it iff there do not exist distinct}} \\
  &~~{\mbox{\it two-cycles $c, c^{\prime} ~{\in}~ H^{2}(M)$ such that}} \\
  &~~{\mbox{\it $w_{2}[c] ~{\neq}~ 0$ and $w_{1} ~{\smile}~ w_{1}[c] = 0$,
but}} \\
  &~~{\mbox{\it $w_{2}[c^{\prime}] ~{\neq}~ 0$ and $w_{1} ~{\smile}~
w_{1}[c^{\prime}]
{}~{\neq}~ 0$}} \\
 & \\
1 &~~{\mbox{\it iff there do exist such two-cycles}} \\
  &~~{\mbox{\it $c, c^{\prime} ~{\in}~ H^{2}(M)$}}
\end{array} \right.}
\]

Now suppose we are given a manifold $M$ with boundary ${\partial}M$. Let
${\tilde M}$ and ${\partial}{\tilde M}$ denote the respective double covers.
Then by Lemma 3, we have
\begin{eqnarray}
u({\partial}M) ~+~ {\chi}(M) =
w_{2}(M) ~+~ w_{1} ~{\smile}~ w_{1}(M)~{\rm mod~2}\; ,
\nonumber \\
 \\
u({\partial}{\tilde M}) ~+~ {\chi}({\tilde M}) =
w_{2}({\tilde M})~{\rm mod~2}\; . \nonumber
\end{eqnarray}
Thus, we see that ${\chi}(M) ~+~ U({\partial}M) =
{\chi}({\tilde M})~{\rm mod~2}$ if
and only if
\[
{w_{2}(M) ~+~ w_{1}(M) ~{\smile}~ w_{1}(M) =
w_{2}({\tilde M})~{\rm mod~2}.}
\]
We therefore obtain
\vspace*{0.3cm}

{\noindent {\bf Lemma 4.} {\it Let $M$ be a smooth non-orientable
four-dimensional manifold with boundary. Let ${\tilde M}$ denote the oriented
double cover of $M$, as above. Then}
\[
{U({\partial}M) ~+~ {\chi}(M) = {\chi}({\tilde M})~{\rm mod~2}}
\]
{\it if and only if ${\delta}(M) = 0$.}}
\vspace*{0.3cm}

{\noindent {\it Proof.} (${\Rightarrow}$) ~Suppose
$U({\partial}M) ~+~ {\chi}(M) = {\chi}({\tilde M})~{\rm mod~2}$,
then $w_{2}(M) + w_{1}(M)
{\smile} w_{1}(M) = w_{2}({\tilde M})~{\rm mod~2}$.
There are three cases to consider:}\vspace*{0.2cm}

{(i) $w_{2}(M) = w_{2}({\tilde M}) = 0$}
\vspace*{0.2cm}

{(ii) $w_{2}(M) = 1$ and $w_{2}({\tilde M}) = 0$}
\vspace*{0.2cm}

{(iii) $w_{2}(M) = 1 = w_{2}({\tilde M})$}\vspace*{0.2cm}

If (i) holds, then $M$ and ${\tilde M}$ are both spin, and so we trivially
have ${\delta}(M) = 0$.

If (ii) holds, then ${\tilde M}$ is spin, but $M$ is not. It follows that
there exists some two-cycle $c ~{\in}~ H^{2}(M)$ such that $w_{2}[c] ~{\neq}~
0$, and that this two-cycle lifts (under ${\pi}:~ {\tilde M}
{}~{\stackrel{\mbox{\scriptsize 2 - 1}} {\longrightarrow}}~ M$) to a two-cycle
${\tilde c} ~{\in}~ H^{2}({\tilde M})$ such that $w_{2}[{\tilde c}] = 0$.
Since $w_{2}[c] ~+~ w_{1} ~{\smile}~ w_{1}[c] = w_{2}[{\tilde c}]$
for any such two-cycle we must have $w_{1} ~{\smile}~ w_{1}[c] = 1$ for all
two-cycles $c ~{\in}~ H^{2}(M)$ such that $w_{2}[c] = 1$. Thus we must have
${\delta}(M) = 0$.

If (iii) holds, then neither $M$ nor ${\tilde M}$ is spin. Since
$U({\partial}M) ~+~ {\chi}(M) = {\chi}({\tilde M})~{\rm mod~2}$, it follows
that
for all two-cycles $c ~{\in}~ H^{2}(M)$ such that $w_{2}[c] ~{\neq}~ 0$, we
must have $w_{1} ~{\smile}~ w_{1}[c] = 0$, and so ${\delta}(M) = 0$.
\vspace*{0.2cm}

(${\Leftarrow}$) ~Conversely, suppose that ${\delta}(M) = 0$. Then case (i)
gives ${\chi}(M) + U({\partial}M) = {\chi}({\tilde M})~{\rm mod~2}$
trivially.

Likewise, if (ii) holds, then on any of the two-cycles, $c$, for which
$w_{2}[c] = 1$, we must have $w_{1} ~{\smile}~ w_{1}[c] = 1$, thus ${\chi}(M)
{}~+~ U({\partial}M) =  {\chi}({\tilde M})~{\rm mod~2}$.

Finally, case (iii) again implies that neither $M$ nor ${\tilde M}$ is spin.
However, ${\delta}(M) = 0$ again implies $w_{2}({\tilde M}) = w_{2}(M) ~+~
w_{1}(M) ~{\smile}~ w_{1}(M)~{\rm mod~2}$ and thus ${\chi}(M) ~+~
U({\partial}M)
= {\chi}({\tilde M}) ~{\rm mod~2}$. \hfill ${\square}$ \vspace*{0.2cm}

Thus, $U({\partial}M) + {\delta}(M) ~{\in}~ {\Bbb Z}_{2}$ is an invariant
which tells us whether the Euler number of a manifold $M$ has the same
parity as the Euler number of the double cover, ${\tilde M}$, of $M$. For
convenience, we will henceforth write
\[
I(M, {\partial}M) = U({\partial}M) + {\delta}(M)\; .
\]

Suppose we are given some four-dimensional manifold $M$ with boundary
${\partial}M$ and its double cover ${\tilde M}$. Next form the
`double' [1] of each manifold, ie. we double $M$ to get $2M$, and ${\tilde
M}$ to get $2{\tilde M}$.
$2{\tilde M}$ is then the oriented cover of $2M$. As
in [1], we have that
\begin{equation}
{{\sigma}(2{\tilde M}) = 0}\; ,
\end{equation}
and so the index of any plane field ${\tilde P}$ on $2{\tilde M}$ becomes
\begin{equation}
{{\mbox{ind}}({\tilde P}, 2{\tilde M}) =
{\mbox{$\frac{1}{4}$}}\bigl(8n - 2{\chi}(2{\tilde M}),\; 8m +
2{\chi}(2{\tilde M})\bigr)}
\end{equation}
where $m, n ~{\in}~ {\Bbb Z}$. We also know that
\begin{equation}
{{\chi}(2{\tilde M}) = 2{\chi}({\tilde M})}\; ,
\end{equation}
and so we obtain
\begin{equation}
{{\mbox{ind}}({\tilde P}, 2{\tilde M}) =
\bigl(2n - {\chi}({\tilde M}),\;  2m + {\chi}({\tilde
M})\bigr)}\; .
\end{equation}

We can assume that there are an equal number of singularities in each
`half' of the double $2M$. As in [1], we then push all of the singularities
over ${\partial}M$ into one of the halves. Then by construction, one of the
halves of $2M$ is free of singularities, and taking this half we have
constructed a non-singular plane field on $M$. The degree of the map from
${\partial}M$ to ${\tilde G_{2, 4}}$ (defined by the plane field) must be,
combining equation (14) with Lemma 4,
\begin{equation}
{(2n - {\chi}(M) + I(M, {\partial}M), 2m +
{\chi}(M) + I(M, {\partial}M))~{\rm mod~2}}
\end{equation}
As in [1], we shall call this degree the {\it Klein kink} of the metric $g_{K}$
(determining the plane field) with respect to ${\partial}M$ and we denote it
\[
{{\mbox{kink}}({\partial}M; g_{K})}\; .
\]
Combining the above, we obtain
\vspace*{0.3cm}

{\noindent {\bf Theorem 2.} {\it Let $M$ be any smooth
four-dimensional manifold with boundary $\partial M \cong \Sigma_{1}
\cup \Sigma_{2} \cup \cdots \cup \Sigma_{n} \neq \emptyset$, where
${\{}{\Sigma}_{i}\; |\;i = 1, \dots, n{\}}$ is some collection of closed
three-manifolds. Then there always exist globally defined non-singular metrics
$g_{K}$ of   Kleinian signature on $M$.  Furthermore every such metric must
satisfy   ${\mbox{kink}}({\partial}M; g_{K}) = (k_{1}, k_{2})$ where }}
\[
{k_{i} = {\chi}(M) ~+~ I(M, {\partial}M)~{\rm mod~2} \hspace*{1cm} i = 1,~2.}
\]
\vspace*{0.3cm}

Thus for an arbitrary non-singular Klein metric $g_{K}$ on $M$,
the parity of the kink number of $g_{K}$ on $\partial M$ is completely
determined.
\vspace*{0.6cm}

\begin{center}
{\bf III. ~Obstructions to pin structures}
\end{center}

As detailed in the Introduction, there are eight double
covers of $O(2, 2)$, which we denote
\[
{h^{a, b, c}:~ {\mbox{Pin}}^{a, b, c}(2, 2) ~{\longrightarrow}~ O(2, 2)}
\]
where $a$ is the sign of the square of space inversion, $b$ is the sign of the
square of time inversion, and $c$ is the sign of the square of the two
combined. The obstructions to constructing a globally well-defined bundle,
with fibre ${\mbox{Pin}}^{a, b, c}(2, 2)$, can be deduced using the
constructions in [3]. Indeed, we obtain
the following:
\vspace*{0.2cm}

{\noindent {\bf Theorem 3.} {\it Let $M$ be a Kleinian four-manifold (with
tangent
bundle ${\tau}(M)$ an $O(2, 2)$ bundle). Then $M$ admits either
${\mbox{Pin}}^{+, +,
+}(2, 2)$ or ${\mbox{Pin}}^{+, +, -}(2, 2)$ structure if and only if}}
\[
{w_{2}(M) = 0}\; .
\]
\vspace*{0.2cm}

{\noindent {\bf Theorem 4.} {\it Let $M$ be a Kleinian four-manifold (with
tangent
bundle ${\tau}(M)$ an $O(2, 2)$ bundle). Then $M$ admits either
${\mbox{Pin}}^{-, +,
+}(2, 2)$ or ${\mbox{Pin}}^{-, +, -}(2, 2)$ structure if and only if}}
\[
{w_{2}(M) ~+~ w_{1}^{+} ~{\smile}~ w_{1}^{+} = 0}\; .
\]
\vspace*{0.2cm}

{\noindent {\bf Theorem 5.} {\it Let $M$ be a Kleinian four-manifold (with
tangent
bundle ${\tau}(M)$ an $O(2, 2)$ bundle). Then $M$ admits either
${\mbox{Pin}}^{+, -,
+}(2, 2)$ or ${\mbox{Pin}}^{+, -, -}(2, 2)$ structure if and only if}}
\[
{w_{2}(M) ~+~ w_{1}^{-} ~{\smile}~ w_{1}^{-} = 0}\; .
\]
\vspace*{0.2cm}

{\noindent {\bf Theorem 6.} {\it Let $M$ be a Kleinian four-manifold (with
tangent
bundle ${\tau}(M)$ an $O(2, 2)$ bundle). Then $M$ admits either
${\mbox{Pin}}^{-, -,
+}(2, 2)$ or ${\mbox{Pin}}^{-, -, -}(2, 2)$ structure if and only if}}
\[
{w_{2}(M) ~+~ w_{1}^{+} ~{\smile}~ w_{2}^{+} ~+~
w_{1}^{-} ~{\smile}~ w_{1}^{-} = 0}\; .
\]
\vspace*{0.2cm}

With these results, we can now investigate the obstructions to pin-Klein
cobordism.
\vspace*{0.6cm}

\begin{center}
{\bf IV. ~Obstructions to pin-Klein cobordism}
\end{center}

In this section ${\{}{\Sigma}_{i}|i = 1, ... n{\}}$ will always denote
some collection of closed three-manifolds.
\vspace*{0.2cm}

{\noindent {\bf Definition:}~ {\it We will say that
there exists a ${\mbox{Pin}}^{a, b, c}(2, 2)$ cobordism for
${\{}{\Sigma}_{i}|i = 1, ... n{\}}$ if and only if there exists a Kleinian
four-manifold $M$ (with a globally non-singular Kleinian metric $g_{K}$)
admitting ${\mbox{Pin}}^{a, b, c}(2, 2)$ structure and satisfying}}
\[
{{\partial}M ~{\cong}~ {\Sigma}_{1} ~{\cup}~ {\Sigma}_{2} ~{\cup}~ ... ~{\cup}~
{\Sigma}_{n}}\; .
\]
\vspace*{0.3cm}

{\noindent {\bf Corollary 1.}
{\it There exists either a ${\mbox{Pin}}^{+, +, +}(2,2)$ or a
${\mbox{Pin}}^{+, +, -}(2, 2)$ cobordism, $M$, for ${\{}{\Sigma}_{i}|i = 1,
... n{\}}$ if and only if}} \[
{(u({\partial}M) ~+~ k_{i} ~+~ I(M, {\partial}M)) = (w_{1}^{+} ~{\smile}~
w_{1}^{+} ~+~
w_{1}^{-} ~{\smile}~ w_{1}^{-})~{\rm mod~2}}
\]
{\it where $k_{i}$ is either of the integers in}
${\mbox{kink}}({\partial}M; g_{K}) = (k_{1}, k_{2})$.
\vspace*{0.3cm}

{\noindent {\it Proof.} (${\Rightarrow}$)~
Suppose such a pin-Klein cobordism, $M$,
exists. Then by Theorem 2, we know that}
\begin{equation}
{k_{i} = ({\chi}(M) ~+~ I(M, {\partial}M))~{\rm mod~2}}
\end{equation}
{(since the Kleinian metric $g_{K}$ is non-singular). Furthermore, by Theorem
3, we must have}
\begin{equation}
{w_{2}(M) = 0}\; ,
\end{equation}
{and by Lemma 3, we know that}
\begin{equation}
{w_{2}(M) ~+~ w_{1}^{+} ~{\smile}~ w_{1}^{+} ~+~ w_{1}^{-} ~{\smile}~ w_{1}^{-}
=
(u({\partial}M) ~+~ {\chi}(M))~{\rm mod~2}}\; .
\end{equation}
{Thus, combining equations (16), (17) and (18), we obtain}
\begin{equation}
{(u({\partial}M) ~+~ k_{i} ~+~ I(M, {\partial}M)) = (w_{1}^{+} ~{\smile}~
w_{1}^{+} ~+~
w_{1}^{-} ~{\smile}~ w_{1}^{-})~{\rm mod~2}}\; .
\end{equation}

{(${\Leftarrow}$) ~Conversely, suppose equation (19) holds.
Take any globally defined
Klein metric $g_{K}$ on $M$, then we must have}
\[
{k_{i} = ({\chi}(M) ~+~ I(M, {\partial}M))~{\rm mod~2}}\; .
\]
Hence $w_{2} = 0$, and so $M$ is pin-Klein with the pin bundle fibre being
${\mbox{Pin}}^{+, +, {\pm}}(2, 2)$. \hfill ${\square}$}
\vspace*{0.3cm}

Using the above proof as a model, we also obtain:
\vspace*{0.3cm}

{\noindent {\bf Corollary 2.} {\it There exists either a ${\mbox{Pin}}^{-, +,
+}(2,
2)$ or a ${\mbox{Pin}}^{-, +, -}(2, 2)$ cobordism, $M$, for ${\{}{\Sigma}_{i}|i
= 1,
... n{\}}$ if and only if}}
\[
{(u({\partial}M) ~+~ k_{i} ~+~ I(M, {\partial}M)) = w_{1}^{-} ~{\smile}~
w_{1}^{-}~{\rm mod~2}}
\]
{\it where $k_{i}$ is either of the integers in}
${\mbox{kink}}({\partial}M; g_{K}) =
(k_{1}, k_{2})$.
\vspace*{0.3cm}

{\noindent {\bf Corollary 3.} {\it There exists either a ${\mbox{Pin}}^{+, -,
+}(2,
2)$ or a ${\mbox{Pin}}^{+, -, -}(2, 2)$ cobordism, $M$, for ${\{}{\Sigma}_{i}|i
= 1,
... n{\}}$ if and only if}}
\[
{(u({\partial}M) ~+~ k_{i} ~+~ I(M, {\partial}M))
= w_{1}^{+} ~{\smile}~ w_{1}^{+}~{\rm mod~2}}
\]
{\it where $k_{i}$ is either of the integers in}
${\mbox{kink}}({\partial}M; g_{K}) =
(k_{1}, k_{2})$.
\vspace*{0.3cm}

{\noindent {\bf Corollary 4.} {\it There exists either a ${\mbox{Pin}}^{-, -,
+}(2,
2)$ or a ${\mbox{Pin}}^{-, -, -}(2, 2)$ cobordism, $M$, for ${\{}{\Sigma}_{i}|i
= 1,
... n{\}}$ if and only if}}
\[
{(u({\partial}M) ~+~ k_{i} ~+~ I(M, {\partial}M)) = 0~{\rm mod~2}}
\]
{\it where $k_{i}$ is either of the integers in}
${\mbox{kink}}({\partial}M; g_{K}) =
(k_{1}, k_{2})$.
\vspace*{0.3cm}

{\noindent Thus we see that the obstructions to ${\mbox{Pin}}^{a, b, c}(2, 2)$
cobordism depend only on boundary data (ie.\ $u({\partial}M)$ and
${\mbox{kink}}({\partial}M; g_{K}) = (k_{1}, k_{2})$),
the values of $a, b ~{\in}~{\{}{\pm}{\}}$, the choice of orientation (ie.\
$w_{1}^{+} ~{\smile}~ w_{1}^{+}$ and $w_{1}^{-} ~{\smile}~ w_{1}^{-}$) and the
invariant $I(M, {\partial}M)$.

Finally, we note that in all the above Corollaries, the expression
$u({\partial}M) + k_{i}
+ I(M, {\partial}M)$ may be replaced by the expression
$u({\partial}{\tilde M}) + k_{i} + {\delta}(M)$, since
$I(M, {\partial}M) = U({\partial}M) + {\delta}(M) = (u({\partial}M) +
u({\partial}{\tilde M})) {\rm mod~2} + {\delta}(M)$.}
\vspace*{0.6cm}

\begin{center}
{\bf V. ~Examples and Applications}
\end{center}

The constuctions introduced in this paper have many applications to
theoretical physics.  We now give some examples and applications.
\vspace*{0.3cm}

{\noindent {\it Example 1.} }  Let $K$ denote the Klein bottle and
$T \cong S^1 \times S^1 $ the Torus.  Then form a Kleinian metric on
$M \cong K \times T$ by taking the product metric formed by using the
natural negative definite metric on $K$ and the natural positive definite
metric on $T$.  Although $M$ is non-time orientable (since traversing the
orientation-reversing loop in $K$ inverts the timelike sub-bundle), $M$ is
still spin since $w_2(K) = 0$, and so all ${\rm Pin}^{a,b,c}(2,2)$ structures
are allowed on $M$.

Now suppose that we take $M$ to be the product space
$M\cong {\Bbb R \Bbb P}^2 \times T$ where, as above, we endow $M$ with the
natural
product metric such that ${\Bbb R \Bbb P}^2$ is timelike.  Then we clearly have
$w_1^- \smile w_1^- ({\Bbb R \Bbb P}^2) = 1$ and $w_2(M) = 1~{\rm mod}~2$.
Thus not all
pin structures will be allowed.  Indeed, one easily calculates that
${\rm Pin}^{+,+,\pm}(2,2)$ and ${\rm Pin}^{-,+,\pm}(2,2)$ structures will not
be allowed, whilst ${\rm Pin}^{+,-,\pm}(2,2)$ and ${\rm Pin}^{-,-,\pm}(2,2)$
are allowed.\vspace*{0.3cm}

{\noindent {\it Example 2.} }  An interesting application is to Kaluza--Klein
type theories in which some of the internal dimensions are allowed to be
timelike.  We
could take the ground state of such a theory to be a manifold of the form
$M\times S^1$ where $M$ is a Lorentzian three-manifold, and the internal space
$S^1$
is timelike.  Then the total metric on $M\times S^1$ would have signature
$--++$, and
the obstruction to this metric being non-singular would again be the condition
that there exists a plane field.  We could even allow the Kleinian metric to
`spin around', so that the internal space fluctuates from being timelike to
being spacelike ( so that the signature of the spacetime $M$ would change from
$-++$ to $+--$).  That is, in terms of the effective three-dimensional theory,
this would correspond to `signature change'.

In general, in order to produce a non-singular theory,
we may wish to consider manifolds $M$ with Kleinian, or even more
exotic, signatures.
For example, if $M \cong S^2\times S^2$ then $M$ does not admit a
non-singular Lorentz metric, but does admit a non-singular Klein metric.
Such choices will generically change the types of Pin structures which are
admitted in the Kaluza--Klein (or other) type theory.\vspace*{0.3cm}

{\noindent {\it Example 3.} }  There has been considerable interest
recently in the study of signature changing spactimes [5--8].
In an extension to the example given in [1] we note that we can have the
nucleation
of a single Kleinian region across a single zero-kink surface homeomorphic to
$S^3$.  As an example, let $M$ be the Kleinian manifold formed by removing
a four-ball from ${\Bbb R \Bbb P}^2 \times S^2$ such that the $S^2$ factor is
timelike.
We have $\partial M \cong S^3$,
and so $I(M, \partial M) = 0$.  In order to produce signature change we require
that the kink on $\partial M$ is zero.  Then since $u(S^3) = 1$, we
see that such a signature change scenario is possible since $M$
admits a ${\rm Pin}^{-,+,\pm}(2,2)$ structure.\vspace*{0.3cm}

{\noindent {\it Example 4.} }  Finally, we note that our results
would be potentially useful in generalising the Penrose flag-plane
construction [16] to
non-orientable manifolds.

\newpage
\begin{center}
{\bf Acknowledgements}
\end{center}

The authors thank Gary Gibbons for many helpful discussions. Thanks also
to Sylvia Fouet for translating most of reference [12] into English, and to
Jo Chamblin for helping prepare this paper. A.C. is supported
by N.S.F. Graduate Fellowship No. RCD-9255644.
\vspace*{0.6cm}

\begin{center}
{\bf References}
\end{center}

{\noindent [1] L. J. Alty and A. Chamblin, {\it Spin Structures on Kleinian
Manifolds}, Class. Quantum Grav. {\bf 11}, 2411--2415 (1994).}\vspace*{0.3cm}

{\noindent [2] L. Dabrowski, {\it Group Actions on Spinors}, Monographs and
Textbooks
in Physical Science, Bibliopolis (1988).}\vspace*{0.3cm}

{\noindent [3] A. Chamblin, {\it On the Obstructions to non-Cliffordian Pin
Structures}, Comm. Math. Phys. {\bf 164}, 65--87 (1994).}\vspace*{0.3cm}

{\noindent [4] F. Wilczek, {\it Fractional Statistics and Anyon
Superconductivity},
World Scientific (1990).}\vspace*{0.3cm}

{\noindent [5] G. Ellis, A. Sumeruk, D. Coule and C. Hellaby,
{\it Change of signature in classical relativity},
Class. Quantum. Grav. {\bf 9} 1535--1554 (1992).}\vspace*{0.3cm}

{\noindent [6] S. A. Hayward, {\it Signature change in general relativity},
Class. Quantum. Grav. {\bf 9} 1851--1862 (1992).}\vspace*{0.3cm}

{\noindent [7] T. Dray, C. A. Manogue and R. W. Tucker, {\it Scalar Field
Equation in the presence of Signature Change}, Phys. Rev. D {\bf 48},
2587--2590
(1993).}\vspace*{0.3cm}

{\noindent [8] L. J. Alty, {\it Kleinian Signature Change},
Class. Quantum Grav. {\bf 11}, 2523--2536 (1994).}\vspace*{0.3cm}

{\noindent [9] J. Barret, G. W. Gibbons, M. J. Perry, C. N. Pope, and P.
Ruback, {\it Kleinian Geometry and the $N = 2$ Superstring},
Int. J. Mod. Phys. A {\bf 9} 1457--1493 (1994). }\vspace*{0.3cm}

{\noindent [10] N. Steenrod, {\it Topology of Fibre Bundles}, Princeton Math.
Series, Princeton Univ. Press (1951).}\vspace*{0.3cm}

{\noindent [11] A. Chamblin, {\it Some Applications of Differential Topology in
General Relativity}, J. Geom. Phys. {\bf 13}, 357--377
(1994).}\vspace*{0.3cm}

{\noindent [12] F. Hirzebruch and H. Hopf, {\it Felder von
fl{\"a}chenelementen in\newline 4-dimensionalen mannigfaltigkeiten},
Math. Ann. {\bf 136}, 156--172 (1958).}\vspace*{0.3cm}

{\noindent [13] E. Thomas, {\it Vector Fields on Manifolds}, Bull. Amer. Math.
Soc. {\bf 75}, 643--668 (1969).}\vspace*{0.3cm}

{\noindent [14] J. W. Milnor and J. D. Stasheff, {\it Characteristic Classes},
Princeton Univ. Press (1974).}\vspace*{0.3cm}

{\noindent [15] M. Kervaire and J. W. Milnor, {\it Groups of Homotopy Spheres:
I}, Ann. of Math. {\bf 77}, (1963).}\vspace*{0.3cm}

{\noindent [16] R. Penrose and W. Rindler, {\it Spinors and Space-time ---
Volume 1},
Cambridge University Press (1986).}\vspace*{0.3cm}

\end{document}